\documentclass[11pt,a4paper]{article}

\usepackage[english]{babel}

\usepackage{amsmath}
\usepackage{amsfonts}
\usepackage{amssymb}
\usepackage{pst-node}

\usepackage{graphicx}

\usepackage{pstricks}
\usepackage{epsfig}

\newcommand{\dparcial}[2]{ \frac{\partial #1}{\partial #2} }

 \author{J. Guerrero$^{1,2}$ \and 
F.F. L\'opez-Ruiz$^{2}$ \and
 V. Aldaya$^{2}$ \and F. Coss\'io$^{2}$}

\title{Harmonic states for the free particle}

\date{\begin{center}
\begin{small}$^1$Departamento de Matem\'atica Aplicada, Universidad de Murcia,
\end{small}\\
\begin{small}Campus de Espinardo, 30100 Murcia, Spain.\end{small}\\
\begin{small}$^2$ Instituto de Astrof\'\i sica de Andaluc\'\i a, CSIC, 
\end{small}\\
\begin{small}
  Apartado Postal 3004, 18080 Granada, Spain
\end{small}\\
\begin{small}juguerre@um.es\ flopez@iaa.es\ valdaya@iaa.es \
fcossiop@gmail.es\end{small}\\                                                  
                                              \end{center}
}

\begin{document}
\maketitle
\begin{abstract}
Different families of states, which are solutions of the time-dependent free
Schr\"odinger equation, are  imported from the harmonic oscillator using the
Quantum Arnold Transformation introduced in \cite{QAT}. Among them, infinite
series of states are given that are normalizable, expand the whole space of
solutions, are spatially multi-localized and are eigenstates of a suitably
defined number operator. Associated with these states new sets of coherent and
squeezed states for the free particle are defined representing traveling,
squeezed, multi-localized wave packets. These states are also constructed in
higher dimensions, leading to the quantum mechanical version of the
Hermite-Gauss and Laguerre-Gauss states of paraxial wave optics. 
Some applications of these new families of states  and  procedures to experimentally
realize and manipulate them are outlined.
\end{abstract}

%

\section{Introduction}

In the context of the quantum free particle, the eigenstates of the
Hamiltonian, which are also eigenstates of the momentum operator,
are not normalizable. These states, the plane waves, are fully
delocalized. However, it is customary to expand any normalizable solution of the
free, time-dependent, Schr\"odinger equation in terms of plane waves, using the
Fourier transform, building in this way wave packets which represent localized
solutions and that are no longer eigenstates of the Hamiltonian. The simplest
examples are the Gaussian wave packets, which have the property of minimizing
the uncertainty relations between the position and the momentum operator.



In \cite{QAT} a transformation was proposed that maps states and operators from
certain quantum systems to the free particle. This transformation is the quantum
version of the Arnold transform (QAT), which in its original classical version
\cite{Arnold} maps solutions of a certain type of classical equation of motion,
a nonhomogeneous linear second order ordinary differential equation (LSODE), to
solutions of the classical equation for the free particle.

In this paper we construct different families of states in the space of
solutions of the quantum free particle  which are the image under the QAT of the
corresponding families in the quantum harmonic oscillator. Among them, we
provide discrete bases of free particle states mapped from the eigenstates of
the harmonic oscillator. They are not eigenstates of the free particle
Hamiltonian, i.e. they are not stationary states, but, rather, eigenfunctions of
a certain quadratic operator $\hat N$ with discrete eigenvalues. The first
states of these bases are Gaussian wave packets, and are localized with
arbitrary initial size,  related to the oscillator frequencies and chosen at
will, and have initial minimal uncertainties. The following ones are
``multi-localized'' in the sense that, for instance, in one dimension the
$n^{\rm th}$-order state presents $n$ zeros and $n+1$ humps which spread out
with time. This mimics the situation for the harmonic oscillator to such an
extent that, in one dimension, it is possible to build creation and
annihilation operators $\hat a^\dagger$ and  $\hat a$. The number operator 
proves to be $\hat N \sim \hat a^\dagger \hat a$, although in the case of the
free particle it is not the Hamiltonian. Going even further, we give a set of
``coherent'' and ``squeezed'' states which are interpreted as traveling,
squeezed wave packets. This construction can easily be generalized to higher
dimensions in different coordinate systems.

These families of states and their construction through the QAT can be
of physical relevance in the description and preparation of initial states, and
even time-evolution processes, in Atom Optics, ion traps physics and
Bose-Einstein Condensates. For instance, the process of turning off and on a
harmonic potential trapping a set of atoms, as well as the production of
coherent and squeezed states in this context, are easily achieved by means of
the QAT. It might also be useful in describing scattering processes in a
discrete basis, instead of the usual approach employing plane waves.


The content of the paper is as follows. Section~\ref{discretebasis-1D} is
devoted to the construction of the discrete bases of states in one dimension by
means of the Quantum Arnold transformation, mapping the eigenstates of the
harmonic oscillator to the free particle Hilbert space.
Section~\ref{estadoscoherentes} deals with the construction of coherent and
squeezed states and the interpretation of them as traveling, squeezed wave
packets. Section~\ref{discretebasis-higherD} generalizes this construction to 2
and 3 dimensions in Cartesian, polar and spherical coordinates, giving rise to
Hermite-Gauss, Laguerre-Gauss and spherical-Gauss wave packets, respectively. 
Section~\ref{physical-applications} presents some possible physical
applications, and proposes experimental settings for producing and manipulating
these states. Section~\ref{Comments} gives some possible generalization of these
constructions and outline possible lines of research. Finally, an Appendix
collects the main formulas for the computation of coherent and squeezed states
for the free particle.


\section{Discrete bases of wave packets in one dimension}
\label{discretebasis-1D}


Let $\cal H$ be the Hilbert space of solutions of the free
particle, time-dependent Schr\"odinger equation, and ${\cal H}_{\rm HO}$ the one
corresponding to the harmonic oscillator of a given frequency $\omega$. 
We shall denote by $\psi(x,t)\in \cal H$
the free particle solutions and by $\psi'(x',t')\in {\cal H}_{\rm HO}$ the
harmonic oscillator ones. Then, the QAT \cite{QAT}, relating solutions
of time-dependent Schr\"odinger equations for the free particle and the harmonic
oscillator, or rather, its inverse, is given by:
\begin{equation}
\psi'(x',t')=
\hat{A}^{-1}\psi(x,t)=\frac{1}{\sqrt{u_{2}(t')}}e^{\frac{i}{2}\frac{m}{\hbar}
                 \frac{1}{W(t')}\frac{\dot{u}_{2}(t')}{u_{2}(t')}
                                                 {x'}^{2}}
\psi(\frac{x'}{u_2(t')},\frac{u_1(t')}{u_2(t')})\,,
\label{qatgeneral}
\end{equation}
where the classical Arnold transformation \cite{Arnold} is explicitly
\begin{equation}
\begin{split}
  A:& \; \; \mathbb R \times T' \longrightarrow \mathbb R \times T\\
    & \quad (x',t')  \longmapsto \; (x,t) =
        A\bigl((x',t')\bigr) = (\tfrac{x'}{u_2(t')},\tfrac{u_1(t')}{u_2(t')})\,.
\end{split}
\end{equation}
$T$ and $T'$ are open intervals of the real line containing $t=0$ and $t'=0$,
respectively, $u_1(t')$ and $u_2(t')$ are two independent solutions of the LSODE
(here dots mean derivation with respect to $t'$):
\begin{equation}
\ddot{x}'+\dot{f}(t')\dot{x}'+\omega(t')^{2}x'=0\,,
\end{equation}
and $W(t')$ is the Wronskian $W(t')=\dot{u}_1(t')u_2(t')-u_1(t')\dot{u}_2(t')$
of the two solutions. For the case of the harmonic oscillator $\dot{f}=0$ and
$\omega(t')=\omega$, and the two independent solutions can be chosen (see
\cite{QAT} for details) as $u_1(t')=\frac{1}{\omega}\sin(\omega t')$
and $u_2(t')=\cos(\omega t')$, with $W(t')=1$. It can be checked that the change
of variables results in\footnote{The change of variables in the particular case
of the harmonic oscillator was already given by Jackiw \cite{Jackiw}.}:
\begin{eqnarray}
 t'&=&\frac{1}{\omega}\tan^{-1}(\omega t)\nonumber\\
 x'&=&\cos(\tan^{-1}(\omega t))x=\frac{x}{\sqrt{1+\omega^2 t^2}}\,.
\label{ArnoldClasico}
\end{eqnarray}

Diagrammatically, the QAT  can be represented as:

\begin{center}
\scalebox{1} 
{
\begin{pspicture}(0,-1.8492187)(7.4490623,1.8492187)
\psline[linewidth=0.02cm,arrowsize=0.05291667cm 2.0,
arrowlength=1.4,arrowinset=0.4]{<-}(1.77,1.1157813)(4.37,1.1157813)
\usefont{T1}{ptm}{m}{n}
\rput(0.9245312,1.1057812){${\cal H}$}
\usefont{T1}{ptm}{m}{n}
\rput(5.5045314,1.0857812){${\cal H}_{\rm HO}$}
\psline[linewidth=0.02cm,arrowsize=0.05291667cm 2.0,
arrowlength=1.4,arrowinset=0.4]{<-}(1.11,0.55578125)(2.59,-1.2842188)
\psline[linewidth=0.02cm,arrowsize=0.05291667cm 2.0,
arrowlength=1.4,arrowinset=0.4]{<-}(4.93,0.55578125)(3.33,-1.2842188)
\usefont{T1}{ptm}{m}{n}
\rput(3.0745313,-1.6142187){${\cal H}_{\rm 0}$}
\usefont{T1}{ptm}{m}{n}
\rput(0.8645313,-0.37421876){$\hat{U}$}
\usefont{T1}{ptm}{m}{n}
\rput(5.204531,-0.35421875){$\hat{U}^{\rm HO}$}
\usefont{T1}{ptm}{m}{n}
\rput(3.2445312,1.6457813){$\hat{A}$}
\end{pspicture} 
}
\end{center}
%
%
%
%
%
where $\hat{U}$ and $\hat{U}_{\rm HO}$ stand for the evolution operators of the
free particle and the harmonic oscillator, respectively, while $\mathcal  H_0$
is the Hilbert space, either for the free particle or the harmonic oscillator,
of solutions of their respective Schr\"odinger equations at $t=0$ (see
\cite{QAT} for the conditions under which $\mathcal  H_0$ is common
to both systems).

Thanks to the commutativity of this diagram, and the unitarity of the operators
appearing in it, we can map objects (wave functions, operators, expectation
values, uncertainties) from one system to the other. In \cite{QAT} we benefited
from this fact for transporting the simplicity of the free particle to more
involved systems, finding, for instance, an analytic expression for the
evolution operator of the complicated systems (even with time-dependent
Hamiltonians) in terms of the free particle evolution operator. In this paper we
shall proceed the other way round, transporting objects and properties from the
harmonic oscillator to the free particle.

Applying now the QAT to the time-dependent eigenstates of the harmonic
oscillator Hamiltonian $\hat H_{\rm HO}'$, with energy
$E_n=\hbar \omega(n+\frac{1}{2})$,
\begin{equation}
\psi'_n(x',t') ={\cal N}_n
\hbox{\Large \it e}^{-i\omega(n+\frac{1}{2}) t'}
\hbox{\Large \it e}^{-\frac{m\omega}{2\hbar}x'^2}
H_n(\sqrt{\frac{m\omega}{\hbar}} x' )\,,
\end{equation}
where ${\cal N}_n=\left(\frac{m \omega }{\hbar \pi}\right)^{\frac{1}{4}
}\frac{1}{\sqrt{2^n n!} }$, we obtain the following set of states, solutions of
the Schr\"odinger equation for the free particle:
\begin{equation}
\psi_n(x,t) ={\cal N}_n\frac{1}{\sqrt{|\delta|}}
\hbox{\Large \it e}^{-\dfrac{x^2\delta^*}{4L^ 2 |\delta|^2}}
\left(\frac{\delta^*}{|\delta|}\right)^{n+\frac{1}{2}}H_n(\frac{x}{\sqrt{2}
L|\delta| } )\,,\label{discretebasis}
\end{equation}
where, in order to obtain a more compact notation, we have introduced the
quantities $L=\sqrt{\frac{\hbar}{2m\omega}}$, with dimensions of length,
and $\tau=\frac{2mL^ 2}{\hbar}=\omega^{-1}$, with dimensions of time. 
We also denote by $\delta$ the complex, time dependent, dimensionless
expression $\delta=1+i\omega t=1+ i\frac{\hbar t}{2mL^2}=1+i \frac{t}{\tau}$.
We have also used the fact that $\hbox{\Large \it e}^{-i\omega t'}=
\hbox{\Large \it e}^{-i {\rm tan}^{-1}(\omega t)}=\frac{\delta^*}{|\delta|}$.
Note that with these definitions the normalization factor ${\cal N}_n$ can be 
written as ${\cal N}_n=\frac{(2\pi)^{-\frac{1}{4}}}{\sqrt{2^n n! L}}$.



This set of states constitutes a basis for the space of solutions of the
free Schr\"odinger equation, since it is mapped from a basis for the harmonic
oscillator through $\hat A$, which is unitary. 


The family of wave functions (\ref{discretebasis}) has been known in the
literature as Hermite-Gauss wave packets \cite{Andrews}, and they have been
widely used, in their two dimensional version (see
Section~\ref{discretebasis-higherD}), in paraxial wave optics 
\cite{Fotonics}\footnote{In \cite{Manko} a basis of discrete states
and their corresponding coherent states were constructed for the damped 
harmonic oscillator, which, in a certain limiting process, would reproduce
ours. Also, in \cite{Suslov-Lopez} the harmonic oscillator wave functions are
mapped to solutions of generalized, time-dependent harmonic oscillator in the
framework of \cite{Suslov-AnPhys}, which resembles the approach in \cite{QAT},
although with somewhat less physical insight.}. However, this kind of states and
the ones constructed in Sections~\ref{estadoscoherentes} and
\ref{discretebasis-higherD} are better understood in the framework of the  QAT.
Note that, making use of the classical solutions only, and through the QAT, we
have been able to import the time evolution from the stationary states of the
harmonic oscillator, $\psi'_n(x',t')$, to the non-stationary ones,
$\psi_n(x,t)$, without solving the time-dependent Schr\"odinger equation.

The first state of this basis, the one mapped from the harmonic oscillator
vacuum state, is given by:

\begin{equation}
\psi_0(x,t) 
   =
\frac{(2\pi)^{-\frac{1}{4}}}{\sqrt{L|\delta|}}
\left(\frac{\delta^*}{|\delta|}\right)^\frac{1}{2}
                    \hbox{\Large \it e}^{-\dfrac{x^2\delta^*}{4L^ 2|\delta|^2}}
		     =\frac{(2\pi)^{-\frac{1}{4}}}{\sqrt{L\delta}}
                     \hbox{\Large \it e}^{-\dfrac{x^2}{4L^ 2 \delta}}\,,
		     \label{paquete-gaussiano}
\end{equation}
which is nothing other than a Gaussian wave packet with center at the origin and
width $L$. The parameter $\tau$ is the dispersion time of the Gaussian wave
packet (see, for instance, \cite{galindo}).
%

\begin{figure*}[h!]
 \centering
   \includegraphics[width=\textwidth]{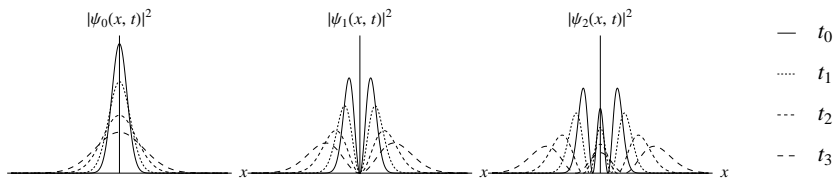}
 \caption{Spreading under time evolution of wave functions $\psi_0$,
$\psi_1$ and $\psi_2$, with $t_k=k\tau$.}
 \label{grafiquita}
\end{figure*}

Figure~\ref{grafiquita} shows some of these wave functions and how they evolve
in time.

We see that the number of ``parts'', or {\it humps}, of the wave functions,
determined by the number of zeros, is ``quantized'', in the sense that there is
only one hump between two consecutive zeros. They are therefore multilocalized,
as the probability of finding the particle is ``peaked'' in separate
regions of space, with nearly zero probability of finding it in between.


The QAT also allows to map invariant operators from one Hilbert space 
(${\cal H}_{\rm HO}$) to the other (${\cal H}$). These invariant operators are
also known as constant or integral of motion operators, in the sense that their
matrix elements are constant with respect to their corresponding time evolution,
and preserve their respective Hilbert spaces. Particularizing the general
expression given in \cite{QAT} or by direct computation, the conserved position
and momentum operator for the harmonic oscillator can be written: 
\begin{align*}
\hat X' &= \frac{\dot{u}_1(t')}{W(t')} x' + \frac{i\hbar}{m} u_1(t')
\frac{\partial}{\partial x'} = \cos \omega t' \, x' + \frac{i\hbar}{m \omega}
\sin \omega t' \frac{\partial}{\partial x'}
\\
\hat P' &= -i \hbar \, u_2(t') \frac{\partial}{\partial x'} - m 
\frac{\dot{u}_2(t')}{W(t')} x' = -i \hbar \cos \omega t' 
\frac{\partial}{\partial x'} + m \omega \,\sin \omega t' \, x' \,,
\end{align*}
and therefore, the conserved creation and annihilation operators are: 
\begin{align*}
\hat a' &= \frac{1}{2L} \hat X' + i \frac{L}{\hbar}\hat P' = 
e^{i\omega t'}\left(\frac{1}{2L}x' + L \frac{\partial}{\partial x'}\right)
\\
\hat {a}'^{\dagger} &= \frac{1}{2L} \hat X' - i \frac{L}{\hbar}\hat P' =
e^{-i\omega t'}\left(\frac{1}{2L} x' - 
L  \frac{\partial}{\partial x'}\right) \,.
\end{align*}
Note that these are the operators acting on solutions of the time-dependent
Schr\"odinger equation for the harmonic oscillator $\psi'_n(x',t')$ as ladder
operators.

Position and momentum operators $\hat{X}'$ and $\hat{P}'$ are mapped into
operators representing conserved position and momentum operators for the free
particle through the QAT:
\begin{align*}
\hat X &= x + \frac{i \hbar}{m} t \frac{\partial}{\partial x} 
\\
\hat P &= -i \hbar \frac{\partial}{\partial x} \,.
\end{align*}%


As a consequence, ladder operators for the harmonic oscillator can be mapped
into ladder operators for the free particle that act as creation and
annihilation operators for the (time-dependent) Hermite-Gauss states:
\begin{eqnarray}
\hat a &=&  L \delta\frac{\partial}{\partial x} +
              \frac{x}{2L}\nonumber\\
\hat a^\dagger &=& -  L \delta^*\frac{\partial}{\partial x} +
              \frac{x}{2L}.   
\end{eqnarray}
The action of $\hat a$ and $\hat a^{\dagger}$ on the Hermite-Gauss wave
functions is the usual one:
\begin{equation}
\hat a  \psi_n(x,t) = \sqrt{n} \,\psi_{n-1}(x,t),  
\qquad 
\hat a^{\dagger} \psi_n(x,t) = \sqrt{n+1}\,\psi_{n+1}(x,t).
\end{equation}
%

It is possible to introduce this discrete basis without resorting to the QAT in
a very intuitive manner. The key point is that the operator $\hat{a}$
annihilates the Gaussian wave packet, and this fact characterizes it.  The whole
family of states (\ref{discretebasis}) can be generated by acting with the
adjoint operator $\hat{a}^\dagger$ of $\hat{a}$. The rest of the construction,
i.e. uncertainties, coherent and squeezed states, etc. would proceed without the
need of resorting to the QAT. However, as we shall see below, the QAT is
very useful when performing involved computations in a very easy way.

In fact, a very useful property of the QAT is its unitarity, implying that
it preserves scalar products and therefore expectation values. This will allow
us to compute expectation values of operators in the free particle in terms of
the corresponding ones in the harmonic oscillator.

Denoting by $\langle\cdot,\cdot\rangle$ the scalar product in the Hilbert space
of solutions of the free particle Schr\"odinger equation, and by 
$\langle\cdot,\cdot\rangle'$ the corresponding one in the harmonic oscillator
Hilbert space, we have that:
\begin{equation}
 \langle\psi,\phi\rangle = \langle\psi',\phi'\rangle'\qquad 
\langle\hat{O}\rangle_{\psi} = \langle\hat{O}'\rangle'_{\psi'}
\label{unitariedad}
\end{equation}
where $\psi,\phi$ and $\hat{O}$ are related to $\psi',\phi'$ and $\hat{O}'$
through the QAT, respectively. It is important to note \cite{QAT} that
(\ref{unitariedad}) is only valid when $\hat{O}$ and $\hat{O}'$ are invariant
operators.
These operators are essentially $\hat X$ and $\hat P$ above, closing a
Heisenberg-Weyl algebra (or $\hat X'$ and $\hat P'$ in the harmonic oscillator),
and their powers. Among them, the most important ones are the quadratic
operators, which together with the Heisenberg-Weyl algebra expand the
Schr\"odinger algebra (see (\ref{chorri})).

%
%

For instance, we can apply this property to compute the uncertainties associated
with each wave function. As a function of time, for each state $\psi_n(x,t)$, they
read:
\begin{equation}
\Delta \hat{x}_n \Delta \hat{p}_n =
(n+\frac{1}{2})\hbar|\delta|=\frac{E_n}{\omega}|\delta|\,.
\label{uncert}
\end{equation}

To compute the uncertainties $\,\Delta
\hat{x}_n=\sqrt{\langle(\hat{x})^2\rangle_{\psi_n}-
 (\langle\hat{x}\rangle_{\psi_n})^2}\,$ and $\,\Delta
\hat{p}_n=\sqrt{\langle(\hat{p})^2\rangle_{\psi_n}-
 (\langle\hat{p}\rangle_{\psi_n})^2}\,$ we have used the transformation of
non-conserved operators when they are written in terms of conserved ones, that
is: 
\begin{eqnarray}
\hat{x} = \hat{X} +\frac{t}{m}\hat{P}&\xrightarrow{ \hat{A}^{-1}}
& 
\hat{X}' +\frac{u_1(t')}{m u_2(t')}\hat{P}' =
\frac{\hat{x}'}{u_2(t')} = |\delta|\hat{x} '\nonumber\\
\hat{p} = \hat{P}&\xrightarrow{ \hat{A}^{-1}}& \hat{P}' 
=\frac{1}{|\delta|}( \hat{p}' +m\omega^2 t\hat{x}')\,, 
\end{eqnarray}
as well as 
\begin{eqnarray}
 \langle(\hat{x})^2\rangle_{\psi_n} &=&
\frac{1}{u_2(t')^2}\langle\left(\hat{x}'\right)^2\rangle'_{\psi'_n} =
|\delta|^2 \frac{\hbar}{m\omega}(n+\frac{1}{2})\nonumber\\
 \langle(\hat{p})^2\rangle_{\psi_n}&=& \langle(\hat{P}')^2\rangle'_{\psi'_n} =
\langle\left(\hat{p}'\right)^2\rangle'_{\psi'_n} =
m \omega\hbar(n+\frac{1}{2})\,,  
\label{uncert2}
\end{eqnarray}
where in the last equality of the equations the
quantum Virial Theorem for the harmonic oscillator potential \cite{galindo},
which is homogeneous of degree two, has been used: 
$\frac{1}{2}m \omega^2 \langle(\hat{x}')^2\rangle'_{\psi'_n} = 
\frac{1}{2m}\langle\left(\hat{p}'\right)^2\rangle'_{\psi'_n} = 
\langle \hat{H}'_{HO} \rangle'_{\psi'_n}$.

%

%
%
For $n=0$ the time evolution of the uncertainty relation (\ref{uncert}) is
the one which results from the usual Gaussian wave packet \cite{galindo}, and,
among all, the minimal one.

The number operator associated with the creation and annihilation operators
above will provide the position of the state in this grid of uncertainties. We
can derive it (or map it from the number operator for the harmonic oscillator)
in the usual way:
\begin{equation}
\hat N = \frac{1}{2}\left( \hat a^\dagger \hat a + \hat a \;\hat
a^\dagger \right)
               =\left[-|\delta|^2L^2\frac{\partial^2}{\partial x^2}
+i\frac{t}{\tau}(x\frac{\partial}{\partial
x}+\frac{1}{2})+\frac{x^2}{4L^2}\right]\,.
\end{equation}
%


By making use of the Schr\"odinger equation, we can turn this operator into a
first order one:
\begin{equation}
     \hat N = \left[i|\delta|^2\tau\frac{\partial}{\partial t}
+i\frac{t}{\tau}(x\frac{\partial}{\partial
x}+\frac{1}{2})+\frac{x^2}{4L^2}\right]\,,
\end{equation}
the validity of which is restricted to solutions of the Schr\"odinger equation.
The action of this operator is such that:
\begin{equation}
     \hat N \psi_n(x,t) = (n+\frac{1}{2}) \psi_n(x,t) =
\frac{E_n}{\hbar\omega}\psi_n(x,t),
\end{equation}
thus reproducing the uncertainties (up to a factor $\hbar$) given in Eq.
(\ref{uncert})
at time $t=0$.

It is quite interesting to note that this operator belongs to the
algebra of the ``maximal kinematical'' symmetry of the free particle, i.e., the
Schr\"odinger group \cite{niederer}, built out of constants of motion operators
up to quadratic order. This symmetry is the standard Galilean symmetry (with
generators $\hat{P}^2=2m\hat H$, $\hat P$, $\hat X$ and the identity $\hat{I}$)
together with spatial dilations (generated by $\hat{XP} \equiv
\frac{1}{2}(\hat{X}\hat{P} + \hat{P}\hat{X})$) and non-relativistic
``conformal'' transformations (with generator $\hat X ^2$). These generators can
be written, in their first-order version, as:
\begin{equation}
\begin{array}{rclrcl}
\hat P &= &- i \hbar \frac{\partial}{\partial x} & \hat P ^2 &= &2m i \hbar
\frac{\partial}{\partial t} \\
\hat X &= &x+  i \hbar\frac{t}{m} \frac{\partial}{\partial x} &
\hat{XP} &= &-2 i\hbar t \frac{\partial}{\partial t} - i\hbar x
\frac{\partial}{\partial x}
-\frac{i\hbar}{2}  \\
 \hat X ^2 &=& 2i\frac{\hbar}{m} t^2 \frac{\partial}{\partial
t} +  2i \frac{\hbar}{m}t x \frac{\partial}{\partial x} + x^2 +
\frac{i\hbar}{m} t\,, & &
\end{array}
\label{chorri}
\end{equation}
providing the non-trivial commutation relations:
\begin{align}
  & & 
  \left[ \hat X,\hat P \right] &= i \hbar & 
  &
\\
  \left[ \hat X, \hat P^2 \right] &= 2 i \hbar \hat P &  
  \left[ \hat X, \hat X^2 \right] &= 0 &
  \left[ \hat X, \hat{XP} \right] &= i \hbar \hat X
\\
  \left[ \hat P, \hat P^2 \right] &= 0 &  
  \left[ \hat P, \hat X^2 \right] &= -2 i \hbar \hat X &
  \left[ \hat P, \hat{XP} \right] &= -i \hbar \hat P 
\\ 
  \left[ \hat X^2 , \hat P^2 \right] &= 4 i \hbar \hat{XP} &  
  \left[ \hat X^2 , \hat{XP} \right] &= 2 i \hbar \hat X^2 &
  \left[ \hat P^2 , \hat{XP} \right] &= -2 i \hbar \hat P^2 
\,.
\end{align}
It is easily checked that $\hat N$ belongs to this Lie algebra, its relation
with the basis above being:
\begin{equation}
\hat N = \frac{1}{\hbar\omega}(\frac{1}{2m}\hat P^2+ \frac{m\omega}{2}\hat X^2) 
         =\frac{1}{\hbar\omega}\hat{H}_{\rm HO}\,,\label{N}
\end{equation}
%
where $\hat{H}_{\rm HO}$ is the operator mapped through the QAT from the
harmonic oscillator Hamiltonian  $\hat{H}_{\rm HO}'$.
See \cite{QAT} for the relevance of quadratic operators like $\hat N$, but
distinct from the Hamiltonian, for building bases of the Hilbert space (see also
\cite{Hioe-Yuen,Manko,Suslov-AnPhys,Cervero}).


%
%

\section{Coherent and squeezed states}
\label{estadoscoherentes}

Once we have successfully imported the eigenstates of the harmonic oscillator, 
we can go further with this scheme and map different families of coherent states
from the harmonic oscillator to the free particle (as we shall see, they might
be defined equivalently, without resorting to the QAT, as eigenstates of the
annihilation operator $\hat{a}$ in the free particle). Even more, squeezed
states may be introduced in a similar fashion. These states have the property
that, having minimal uncertainty relations, the uncertainty in position (or
momentum) is squeezed, at the expense of increasing the uncertainty in momentum
(or position). These states are used in every day experiment in Quantum Optics
since they have low noise and allow to measure quantities with higher precision
\cite{CitaSqueezed}.  

The most general squeezed and displaced number states $|n,\xi,a\rangle'$ for
the harmonic oscillator are built using
the squeezing operador
$\hat{S}'(\xi)=e^{\frac{1}{2}(\xi^* \hat{a}'^2-\xi (\hat{a}'^\dag)^2)}$ and the
displacement operator $\hat{D}'(a)=e^{a \hat{a}'^\dag-a^*\hat{a}'}$:
\begin{equation}
 |n,\xi,a\rangle' =\hat{D}'(a)\hat{S}'(\xi)|n\rangle'
\end{equation}
%
%
where $a$ is, as usual, the complex number
\begin{equation}
a=\sqrt{\frac{m\omega}{2\hbar}}x_0+i\frac{1}{\sqrt{2m\hbar\omega}}p_0 =
\frac{x_0}{2 L}+i\frac{p_0 L}{\hbar}=\frac{1}{2L}(x_0+iv_0\tau)\,, \label{a}
\end{equation}
with $v_0=\frac{p_0}{m}$, and $\xi$ is the squeezing parameter that will be
taken real, $\xi=r$, for simplicity\footnote{The difference between real and imaginary squeezing parameter is easily seen checking the action of the squeezing operator on the Wigner quasi-probability distribution, with domain the phase space $(x,p)$. Real $\xi$ implies squeezing in the $x$ or $p$ direction, while imaginary $\xi$ squeezes in the diagonal $x+p$ or $x-p$ directions. In other words, when $\xi= \rho e^{i \theta}$, the squeezing 
takes place in the direction $\frac{\theta}{2}$ in the plane $(x,p)$.}.

The complete expression for the configuration space wave functions of
these states for the harmonic oscillator, and its free particle version through
the QAT, is given in the Appendix. Here we only give their general
form for the free particle:  
\begin{eqnarray}
 \psi_{(a,r)}^n(x,t)=  {\cal N}_n e^{r/2}\frac{1}{\sqrt{|\delta_r|}}
\left(\frac{\delta_r{}^*}{|\delta_r|}\right)^{n+\frac{1}{2}}
 \hbox{\Large \it e}^{i\omega t \frac{x^2}{4L^2|\delta|^2}}
\hbox{\Large \it e}^{i\theta(x,t)}\hbox{\Large \it e}^{-q^2/2} H_n(q)\,,
\label{squeezedstates-freeparticle-text}
\end{eqnarray}
where $q= \frac{x -x_0+\frac{p_0}{m} t}{\sqrt{2}Le^{-r}|\delta_r|}$ and 
$\delta_r=1+i e^{2r}\omega t$, and the phase $\theta(x,t)$ is given in 
Eq. (\ref{free-theta}) in the Appendix. 

Now we focus ourselves on some particular cases, giving their physical
interpretation in the free particle system. 
By putting $n=r=0$ in (\ref{squeezedstates-freeparticle-text}), we obtain
coherent states for the free particle:
\begin{equation}
\phi_{a} (x,t) = \frac{(2\pi)^{-\frac{1}{4}}}{\sqrt{L|\delta|}}
\left(\frac{\delta^*}{|\delta|}\right)^\frac{1}{2}
             \hbox{\Large \it e}^{-\dfrac{(x-x_0)^2+x_0v_0 t+i\tau v_0(v_0
t-2x+x_0)}{4L^ 2 \delta}}\,,
 \label{estadocoherente}
\end{equation}
which could also have been obtained as eigenstates of the annihilation operator 
since they verify:
\begin{equation}\hat a \, \phi_{a} (x,t) = a \,  \phi_{a} (x,t)\,.
\end{equation}

These states can also be obtained by the action of the imported displacement 
operator $\hat{D}(a)=e^{a \hat{a}^\dag-a^*\hat{a}} =
e^{\frac{i}{\hbar}(p_0\hat{X}-x_0\hat{P})}$, which is, up to a phase, a Galilean
boost with parameter $p_0$ together with a translation of parameter $x_0$ on the
Gaussian wave packet. They constitute an over-complete set of the Hilbert space
of the free particle and represent traveling Gaussian wave packets, with mean
momentum and initial position  $p_0$ and $x_0$, respectively. These states are
not eigenstates of the number operator $\hat{N}$. The expectation values of this
operator on these coherent states can be computed importing the result from the
harmonic oscillator using (\ref{unitariedad}) and (\ref{N}):
\begin{equation}
\langle \phi_{a} |\hat{N}|\phi_{a} \rangle = \hbar(|a|^2 +\frac{1}{2})\,.
\end{equation}

Setting now $n\neq 0$, $r=0$ in (\ref{squeezedstates-freeparticle-text}), we
obtain a family of states of the free particle that could be obtained by acting
with Galilean boosts and translations on a fixed state of the basis, 
$\psi_n (x,t)$, and constitutes a new over-complete set of states for the free
particle Hilbert space:
\begin{eqnarray}
\phi_{a}^n (x,t) &=&  {\cal N}_n \frac{1}{\sqrt{|\delta|}}
\left(\frac{\delta^*}{|\delta|}\right)^{n+\frac{1}{2}}H_n(\frac{x-x_0-v_0t}{
\sqrt { 2 }
L|\delta| } )\nonumber\\
& & \hbox{\Large \it e}^{-\dfrac{(x-x_0)^2+x_0v_0 t+i\tau v_0(v_0
t-2x+x_0)}{4L^ 2 \delta}},
\label{estadocoherenteN}
\end{eqnarray}
%
representing traveling multi-localized wave packets bearing $n+1$ humps, with
mean momentum and initial position  $p_0$ and $x_0$, respectively, where $a$ is
given by (\ref{a}).
As in the case $n=0$, these states are not eigenstates of the number
operator, but the expectation values, computed as before, are:
\begin{equation}
\langle \phi_{a}^n |\hat{N}|\phi_{a}^n \rangle = \hbar(|a|^2 +n+\frac{1}{2})
\,
. \label{numberexpectation}
\end{equation}
Being a set of coherent states, the uncertainty relations of $\psi_a^n,\,\forall
a\in \mathbb{C}$, are the same as those of $\psi_n$ given in Eq. (\ref{uncert})
(see \cite{perelomov}).

Finally, let us discuss the simplest case of squeezed state, that with 
$r\neq 0$, $n=0$ and $a=0$, corresponding to the squeezed vacuum in the 
harmonic oscillator or a ``squeezed'' Gaussian wave packet in the free
particle:

\begin{equation}
\varphi_r(x,t)= 
\frac{(2\pi)^{-\frac{1}{4}}}{\sqrt{L|\delta_r|}}e^{r/2}
\left(\frac{\delta_r^*}{|\delta_r|}\right)^\frac{1}{2}
                     \hbox{\Large \it e}^{-\dfrac{x^2e^{2r}\delta_r^*}{4L^ 2 |\delta_r|^2}} \,.
		     \label{paquete-gaussiano-squeezado}
\end{equation}

This wave function corresponds to a Gaussian wave packet where the position
and time variables have been rescaled, in addition to the wave function, as
\begin{equation}
 \varphi_r (x,t) = e^{r/2} \psi_0(e^{r}x,e^{2r}t)\,.
\end{equation}

Again, this can be interpreted as the action of the imported displacement 
operator $\hat{S}(r)=e^{\frac{r}{2}( \hat{a}^2-(\hat{a}^\dag)^2)}=
e^{\frac{i r}{\hbar}\hat{XP}} $ which turns out to be a dilation.

Comparing the expressions of (\ref{paquete-gaussiano}) and
(\ref{paquete-gaussiano-squeezado}), we realize that 
(\ref{paquete-gaussiano-squeezado}) corresponds to a Gaussian wave packet with
initial width $L_r=e^{-r}L$. This can be easily understood since in the free
particle there is no natural frequency, and therefore there is no natural width
$L$, so that squeezing can be absorbed in the width $L$ of the wave packet.

\section{Discrete bases of wave packets in higher dimensions}
\label{discretebasis-higherD}

The generalization of the previous construction to higher dimensions is
immediate. In more dimensions, due to the symmetries of the harmonic oscillator,
we can choose different separation of variables to solve the Schr\"odinger
equation for the harmonic oscillator, which amounts to find simultaneous
eigenstates for the harmonic oscillator Hamiltonian and other operators
commuting with it and among themselves. For instance, separation of variables in
cartessian coordinates in N-dimensions is equivalent to diagonalizing
simultaneously the harmonic oscillator Hamiltonian, and the 1D harmonic
oscillator in each coordinate. This can always be done even if the
frequencies for each direction are different. The common basis of eigenstates is
the product of 1D harmonic oscillator eigenstates in each variable.
If we impose spherical symmetry, then all frequencies must coincide and we can
search for common eigenstates of the Hamiltonian and the angular momentum
operators commuting with it and among themselves (for instance, in three
dimensions they would be $\hat{L}^2$ and $\hat{L}_z$, and in two dimensions it
would be only $\hat{L}$).

For the case of Cartesian coordinates, the generalization of the previous
sections is immediate by a direct generalization of the QAT to higher
dimensions, the discrete basis of states for the N-dimension free particle being
products of N copies of (\ref{discretebasis}) in each of the $N$ variables
$(x_1,x_2,\ldots,x_N)$. If $\psi_{n_i}(x_i,t)$ represents a wave packet of the
form given in (\ref{discretebasis}) but in the variable $x_i$ and with width
$L_i$ (and with the corresponding frequency $\omega_i$), then the
$N$-dimensional wave packets are:
\begin{equation}
\psi_{\vec{n}}(\vec{x},t) =
\Pi_{i=1}^N \psi_{n_i}(x_i,t)\,,\label{discretebasisND}
\end{equation}
where $\vec{n}=(n_1,n_2,\ldots,n_N)$. If all the widths $L_i=L$ are identical,
it can be written as:
\begin{equation}
\psi_{\vec{n}}(\vec{x},t) =\frac{(2\pi)^{-\frac{N}{4}}}{(L|\delta|)^{N/2}}
\hbox{\Large \it e}^{-\dfrac{\sum_{i=1}^N x_i^2}{4L^ 2 \delta}}
\left(\frac{\delta^*}{|\delta|}\right)^{\sum_{i=1}^N n_i+\frac{N}{2}}\Pi_{i=1}^N
\frac{1}{\sqrt{2^{n_i} n_i!}}H_{n_i}(\frac{x_i}{\sqrt{2}
L|\delta| } )\,,\label{discretebasisND2}
\end{equation}

Coherent states, or traveling wave packets, are defined similarly, as products
of $N$ copies of (\ref{estadocoherenteN}), with a vector 
$\vec{a}\in \mathbb{C}^N$:
\begin{equation}
\psi_{\vec{a}}^{\vec{n}}(\vec{x},t)=\Pi_{i=1}^N \psi_{a_i}^{n_i}(x_i,t)\,.
\end{equation}

For $N=2$ these states have been widely used in the paraxial approximation to
the Helmholtz equation of wave optics, known as the Hermite-Gauss states
\cite{Fotonics}. Due to the analogy of the paraxial approximation to the
Helmholtz equation (with the $z$ coordinate acting as time) and Schr\"odinger 
equation in two dimensions, they have also been exploited in atom optics
and matter waves \cite{Borde}. 

\begin{figure*}
 \centering
   \includegraphics[height=5cm]{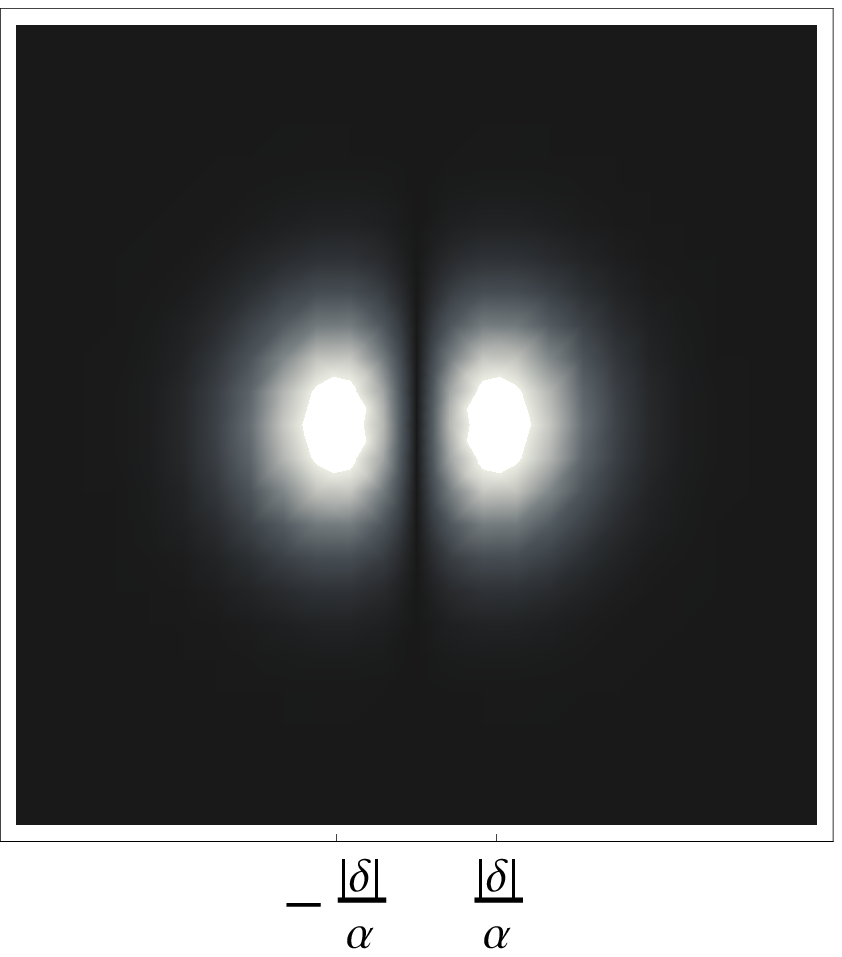}\hspace{1cm}
\includegraphics[height=5cm]{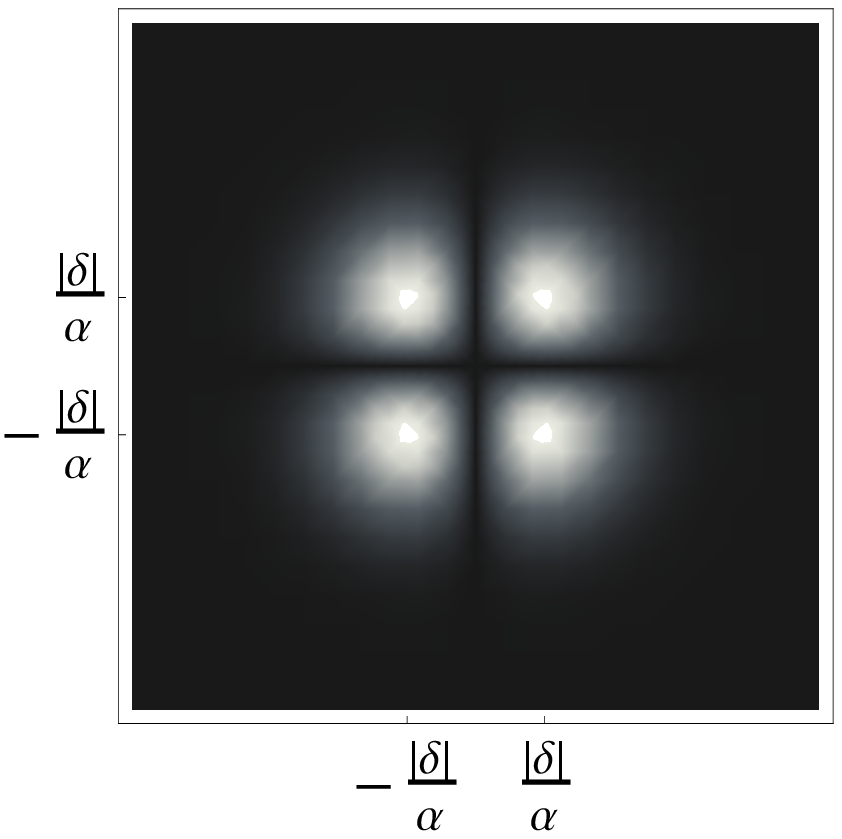}
 \caption{Density plot
of the probability distribution for the Hermite-Gauss states $\psi_{(1,0)}$ and
$\psi_{(1,1)}$. The positions of the maxima are shown.}
 \label{grafiquita2Dc}
\end{figure*}

In Figure \ref{grafiquita2Dc} a density plot of the probability distribution for
the Hermite-Gauss states $\psi_{(1,0)}$ and $\psi_{(1,1)}$ is shown.

In the case of cylindrical symmetry in paraxial wave optics, the polar version
of these estates have been used, known as Laguerre-Gauss states
\cite{Fotonics}.

The corresponding states for the free Schr\"odinger equation in two dimensions,
derived throught the QAT from the 2D harmonic oscillator in polar coordinates,
are:
\begin{eqnarray}
 \psi_{n,l}^{\pm}(r,\phi,t)&=&\sqrt{\frac{n!}{2\pi \Gamma(n+l+1)L^2|\delta|}}
\left(\frac{\delta^*}{|\delta|}\right)^{2n+l+1}\hbox{\Large \it e}^{\pm i l
\phi}
\hbox{\Large \it e}^{-\frac{r^2}{4L^2\delta}}\nonumber \\
& & \left(\frac{r}{\sqrt{2}L|\delta|}\right)^l
L_n^l(\frac{r^2}{2L^2|\delta|^2})\,,\label{laguerre}
\end{eqnarray}
where $n,l=0,1,2\ldots$, and $L_n^l(x)$ are Laguerre polynomials. The state with
$n=0,l=0$ is the Gaussian wave 
packet in two dimensions.

\begin{figure*}
 \centering
   \includegraphics[height=5cm]{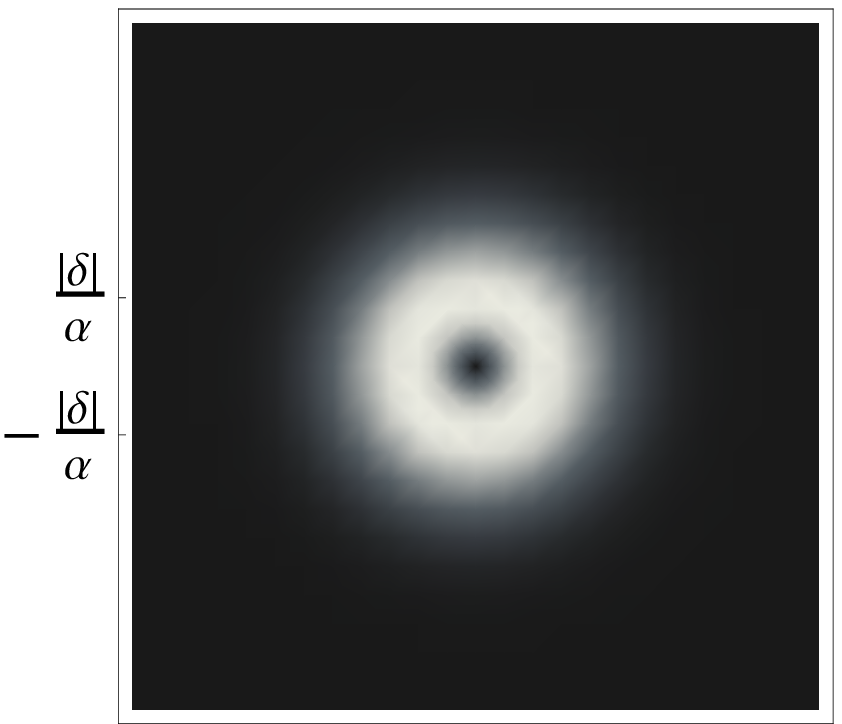}\hspace{0.1cm}
\includegraphics[height=5cm]{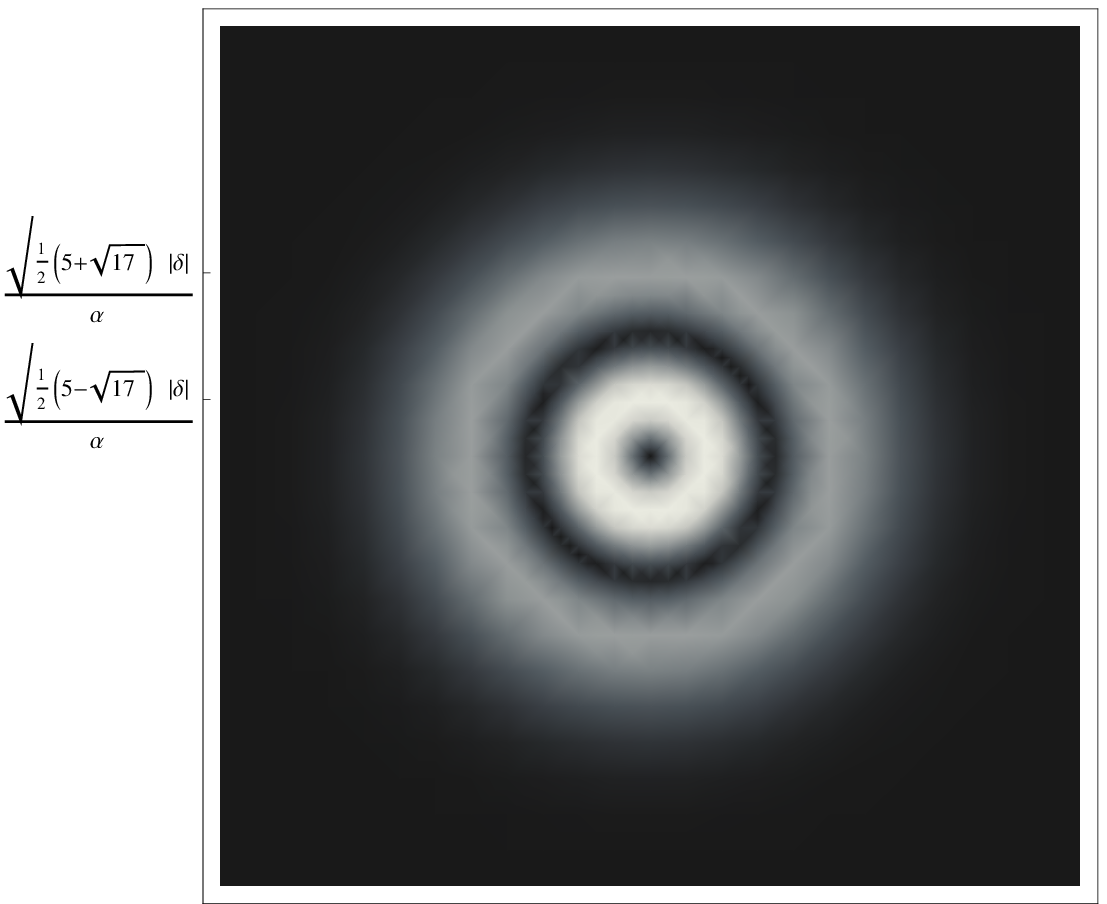}
 \caption{Density plot
of the probability distribution for the Laguerre-Gauss states $\psi_{0,1}^{\pm}$ and
$\psi_{1,1}^{\pm}$. The positions of the maxima are shown.}
 \label{grafiquita2Dp}
\end{figure*}

These states are eigenstates of the angular momentum operator $\hat{L}$ in 
2 dimensions, with values $\hat{L}\psi_{n,l}^{\pm}(r,\phi,t)=\pm l
\psi_{n,l}^{\pm}(r,\phi,t)$. In Figure \ref{grafiquita2Dp} density plots
of the probability distribution for the Laguerre-Gauss states $\psi_{0,1}^{\pm}$ and
$\psi_{1,1}^{\pm}$ are shown.

The generalization to three dimensions in spherical coordinates is
straightforward, the states having the form of (\ref{laguerre}) but in terms of
spherical harmonics and confluent hypergeometric functions
$M(\cdot,\cdot\:;\cdot)$  (which are also polynomials in this case):
\begin{eqnarray}
 \psi_{n,l,m}(r,\theta,\phi,t)&=&\sqrt{\frac{\Gamma(l+3/2+n-1)}{\sqrt{2}
(n-1)!\Gamma(l+3/2)^2L^3|\delta|}}
\left(\frac{\delta^*}{|\delta|}\right)^{2(n-1)+l+3/2}
\hbox{\Large \it e}^{-\frac{r^2}{4L^2\delta}}\nonumber \\
& &Y_l^m(\theta,\phi) \left(\frac{r}{\sqrt{2}L|\delta|}\right)^l
M(-n+1,l+3/2;\frac{r^2}{2L^2|\delta|^2})\,,\label{3D}
\end{eqnarray}

None of these states are eigenstates of the free particle Hamiltonian, but can
be seen to have expectation values of the energy equal to half the energy of the
corresponding, through the QAT, harmonic oscillator eigenstates\footnote{This a
consequence, again, of (\ref{unitariedad}) and the quantum Virial theorem for
the quadratic, homogeneous harmonic oscillator potential.}:
\begin{equation}
\begin{array}{rclrcl}
\langle \hat{H}\rangle_{\psi_{\vec{n}}}&=& \frac{1}{2} E_{\vec{n}}\quad &
E_{\vec{n}} &=& \hbar\omega (\sum_{i=1}^N n_i+\frac{N}{2})\\
\langle \hat{H}\rangle_{\psi_{n,l}^\pm}&=& \frac{1}{2} E_{nl}\quad    &E_{nl}&=&
\hbar\omega(2n+l+1)\\
\langle \hat{H}\rangle_{\psi_{n,l,m}}&=& \frac{1}{2}E_{nlm}\quad & E_{nlm} &=&
\hbar\omega(2(n-1)+l+\frac{3}{2})
\end{array}
\end{equation}

These states, as in the one dimensional case, have many humps. For Cartesian
coordinates they have $\Pi_{i=1}^N (n_i+1)$ humps, while in polar coordinates
they have $(n+1)$ humps (seen in the radial coordinate) with annular form.

Uncertainty relations can also be computed for these states. However, only the
cartessian version of the operators $\vec{x}$ and $\vec{p}$ can
be used since the canonical momentum associated with the radial coordinate, $p_r$,
is not self-adjoint. The expressions, computed as in the
one dimensional case, are:
\begin{eqnarray}
 (\Delta x_i)_{\psi_{\vec{n}}} (\Delta p_i)_{\psi_{\vec{n}}}
&=&\frac{1}{\omega}|\delta| E_{n_i} \,,\quad i=1,\ldots N\nonumber\\
(\Delta x)_{\psi_{nl}} (\Delta p_x)_{\psi_{nl}} &=&(\Delta y)_{\psi_{nl}}
(\Delta p_y)_{\psi_{nl}} =\frac{1}{2\omega}|\delta| E_{nl} \\
(\Delta x)_{\psi_{nlm}} (\Delta p_x)_{\psi_{nlm}} &=&(\Delta y)_{\psi_{nlm}}
(\Delta p_y)_{\psi_{nlm}}
=(\Delta z)_{\psi_{nlm}} (\Delta p_z)_{\psi_{nlm}}\nonumber\\ 
&=&\frac{1}{3\omega}|\delta| E_{nlm} \nonumber
\end{eqnarray}

As in the one dimensional case,  squeezed states, interpreted as rescaled
wave packets, can also be introduced in higher dimensions.

\section{Physical applications}
\label{physical-applications}

The theoretical relevance of these multi-localized, squeezed, traveling wave packets 
for the free particle, as well as their
construction by means of the QAT, is of no doubt.  
From a more practical point of view, 
the preparation of this kind of discretized free states might be achieved by the
use of a harmonic oscillator the potential of which is switched off at a given
time\footnote{This idea was proposed in \cite{QuantumSling}, and was named
``Quamtum Sling''.}. The vacuum state of this harmonic oscillator, when switched
off, will provide the ``vacuum'' Gaussian wave packet with width
$L=\sqrt{\frac{\hbar}{2m\omega}}$, where $m$ is the mass of the particle and
$\omega$ the frequency of the oscillator. Note that the dispersion time $\tau$ 
coincides with the inverse of the frequency of the oscillator. If the harmonic
oscillator is in the $n$-th excited state, the $(n+1)$-hump state is obtained.
To obtain traveling states, the initial state should be a coherent state
$\phi_a(x,t)$ of eq. (\ref{estadocoherente}) for a one-hump traveling state or
$\phi_a^n (x,t)$ for a $(n+1)$-hump traveling state. These coherent states can
be obtained by acting with time-dependent classical forces on the harmonic
oscillator according to Glauber \cite{Glauber} (see also
\cite{perelomov,QuantumNoise}). In fact, if the classical force is given by the
potential $V(x)=- f(t) x$, and the initial state is the vacuum $|0\rangle$,
then, if the force $f(t)$ has finite duration or it is fast decaying in time,
after a suitable time a standard coherent state $|a\rangle$ is obtained with
$a=\frac{i}{\sqrt{\omega/\pi}}\hat{f}(\omega)^*$, where $\hat{f}(\omega)$ is the
Fourier component of $f(t)$ in the frequency $\omega$ of the oscillator. For an
arbitrary force $f(t)$ a time-dependent coherent state $|a(t)\rangle$ is
obtained, see \cite{perelomov,QuantumNoise}.




It is possible to ``capture'' one of these traveling states at a later time
$t_1$ switching on a harmonic oscillator potential with an appropriate 
frequency $\omega_1$ that would ``freeze'' and keep it in an coherent state
(without dispersion), until the potential is switched off again at a time $t_2$
and the wave packet is released, traveling again as a free wave packet that
disperses in time. This way, information can be stored temporally in this
``oscillator traps'', which can also be used to further manipulating them or
even measuring the resulting state by means of adequate lasers. The frequency
$\omega_1$ required to capture the dispersed wave packet should be fine tuned in
such a way that the wave packet, at the time $t_1$, matches an appropriate
solution (with the same $n$ and the same $a$) of the harmonic oscillator with
frequency $\omega_1$. Using the relation between $\omega$ and $L$ (valid at
$t=0$), $\omega= \frac{\hbar}{2 m L^2}$, and the fact that the width of the wave
packet scales as $|\delta|$, see eq. (\ref{uncert2}), then $\omega_1=
\frac{\hbar}{2 m L^2|\delta_1|^2}$, where $\delta_1=1+i\omega t_1$. This
expression can be written in terms of distances using the mean velocity $v_0$ of
the wave packet encoded in the complex number $a$.

This process can be elegantly described by a sequence of QAT and evolution 
operators in the following way. Denote by $\hat{A}_{\omega,t_0}(t)$ the QAT from a
harmonic oscillator of frequency $\omega$ performed at time $t$, and by
$\hat{U}'_\omega$ the unitary time evolution operator for that harmonic
oscillator\footnote{With  this notation, for $t = t_0$, $\hat{A}_{\omega,t_0}(t_0)$ is the identity (classical solutions $u_1(t')$ and $u_2(t')$ have been appropiately chosen, see  \cite{QAT}), and $t=\frac{u_1(t')}{u_2(t')}$.}. Then the state after the process described above is (only the time dependence is explicit):
\begin{equation}
\begin{split}
\psi(t) 
&=\hat{U}(t,t_2)\hat{U}'_{\omega_1}(t_2,t_1) \hat{U}(t_1,t_0)\psi'(t_0) =
\\
&= \hat{A}_{\omega_1,t_2}(t) \hat{U}'_{\omega_1}(t',t_2)
\hat{U}'_{\omega_1}(t_2,t_1)
\hat{A}_{\omega,t_0}(t_1) \hat{U}'_{\omega}(t'_1,t_0)\psi'(t_0) =
\\
&=\hat{A}_{\omega_1,t_2}(t) \hat{U}'_{\omega_1}(t',t_1)
\hat{A}_{\omega,t_0}(t_1) \hat{U}'_{\omega}(t'_1,t_0)\psi'(t_0)\,,
\end{split}
\end{equation}
%
%
%
where $\psi'(t_0)$ is the starting harmonic oscillator state. This way, in 
the last expression only evolution operators for harmonic oscillators appear. 

Note that if the frequency of the harmonic oscillator is not modified,
and the same frequency  $\omega$ is used to capture the state in the harmonic
oscillator trap, the resulting state will be a squeezed state with squeezing
parameter $r$ given by $r=-\log(|\delta_1|)$, which is negative.
This can be seen as a feasible way of producing squeezing in trapped states, 
simply switching off-switching on the trap for a period of time $t$, resulting
in a squeezing parameter $r=-\frac{1}{2}\log(1+\omega^2 t^2)$.
In fact, a similar way of producing squeezed states in Bose-Einstein 
Condensates was reported  in \cite{bec1}.

The states here constructed are rather robust in the sense that 
the number of humps is conserved even in the presence of small
perturbations. Numerical calculations in one dimension have been performed,
simulating perturbations by square potentials (well or barriers),
leading to the conclusion that this holds as long as the mean
energy of the state is large compared with the scale of the perturbing 
potential and the wave packet is sharp enough in momentum space in such a
way that the transmission coefficient can be considered a
constant. Under these circumstances (see for instance
\cite{amjphys}), the wave packet behaves as a plane wave and the
effect of the barrier in the transmitted packet is  an overall
attenuation, preserving its shape, and a time delay which takes
its maximum values for energies near the resonant ones (and where
the transmission coefficient is one). As shown in \cite{amjphys},
this result is valid for any bounded potential of compact support,
provided that the width of the potential is small in the sense
that the time to pass through the barrier is smaller than the
dispersion time of the wave packet $\tau$.
Therefore, the conclusions obtained with the square potential can
be generalized to any finite-range bounded potential.

These wave packets are also robust under the influence of time dependent, 
homogeneous external fields, or even linear damping. In these cases, the
centroid of the wave packets follows the classical trajectories, but their
shapes are unaltered, apart from the unavoidable
dispersion\footnote{Regarding the dispersion effect, which should be taken
into account for practical application in Atomic Physics or BEC, it takes place
in a time scale given by the dispersion time $\tau$.  For instance, for a
condensate of Na atoms with size $\approx 1 mm$,  this time scale is $\approx
500 s$. } 
\cite{Andrews}. The case of linear damping is interesting due to the fact that
the presence of damping prevents the dispersion of the wave packets, which
asymptotically have finite width \cite{Manko,Andrews}.
The case of homogeneous external fields is particularly interesting, since
it includes the case of free fall, which is a common experimental situation in
atomic physics or Bose-Einstein Condensates (see for instance \cite{BEC-fall}).

Under the conditions commented above, these wave packets
evolve without distortion even in the presence of perturbations. However,
one could be interested, acting with appropriate potentials, in obtaining 
transitions between wave packets with different number of humps, 
in such a way that, for instance,  a one-hump packet splits into a 
two-hump packet or a two-hump packet coalesces into a one-hump packet. This
would open the door to performing quantum gates acting on q-bits realized with
the one-hump and the two-hump states. A way of implementing this is to benefit
from the fact that wave packet dynamics is similar to wave optics in the sense
that an analogue to the ABCD law for optics is satisfied for wave packets
\cite{Borde2}. Even the transmutation of Hermite-Gauss wave packets into
Laguerre-Gauss wave packets can be achieved using ABCD matrices, implementing
a mode converter \cite{modeconverter}.

Among 
further theoretical applications, we could think of
expanding plane waves in terms of the discrete basis
$\{\phi_{\vec{a}}^{\vec{n}}\}_{\vec{n}\in\mathbb{N}_o^N}$, with fixed
$\vec{a}\in\mathbb{C}^N$, and describing scattering process in a discrete basis,
or expanding arbitrary wave packets in a continuum over-complete set
$\{\phi_{\vec{a}}^{\vec{n}}\}_{\vec{a}\in \mathbb{C}^N}$, with fixed
$\vec{n}\in\mathbb{N}_o^N$, which could be discretized in a lattice
$\mathbb{Z}^N\times\mathbb{Z}^N$ of points to perform numerical computations,
while keeping the over-complete character. Note that they are over-complete
for $t=0$ if the volume of the unit cell is smaller or equal to $\hbar^N$,
and again by the unitary time evolution they continue to be over-complete for
any $t$, see \cite{perelomov}. Similar constructions, like
the Harmonic Oscillator (HO) method \cite{Moshinsky} or the Transformed Harmonic
Oscillator (THO) method \cite{Petkov} have been proposed, mainly in nuclear
physics, to describe the bound states and the continuum spectrum in a discrete
basis. But there the construction just appears as a mathematical tool for
approximating the solutions, with no physical meaning. Our states, however, are
physically meaningful (as traveling wave packets) and experimentally feasible.

\section{Comments and outlook}
\label{Comments}


In this paper we have constructed different families of free particle 
states mapping harmonic oscillators states through the QAT. It would also be 
possible to generalize the construction to other important families of harmonic
oscillator states, \textit{Gaussian states}, which are mixed states with the 
property that their Wigner distribution functions are Gaussian (see, for 
instance, \cite{QuantumNoise}). They include coherent states, squeezed states 
and thermal equilibrium states. Their main property is that Gaussian states 
keep this property under evolution even in the presence of dissipation and 
decoherence. That is the reason why they have been profusely used in Quantum 
Optics and Quantum Information in the Continuous Variables formalism 
\cite{GaussianStates,Cirac}.

We should stress that, by using the QAT,  the density matrix $\hat{\rho}'$ 
of a mixed harmonic oscillator state can be mapped into the density matrix
$\hat{\rho}$ of a mixed free particle state:
\begin{equation}
 \hat{\rho} = \hat{A}\hat{\rho}' \hat{A}^{\dag}
\end{equation}

The unitarity of $\hat{A}$ guaranties that $\hat{\rho}$ is a proper density
matrix, provided that $\hat{\rho}_{\rm HO}$ is. Even more, it is easily
checked using the evolution operator constructed in \cite{QAT} that:
\begin{equation}
 \dparcial{\hat{\rho}'}{t} = -\frac{i}{\hbar} [\hat{H}_{\rm HO}, \hat{\rho}'] \quad
 \Rightarrow \quad \dparcial{\hat{\rho}}{t} = -\frac{i}{\hbar} [\hat{H}, \hat{\rho}]
\end{equation}
i.e., if $\hat{\rho}'$ satisfies the quantum Liouville equation for the
harmonic oscillator Hamiltonian, then $\hat{\rho}$ satisfies the free particle
counterpart.

All the properties of the density matrix $\hat{\rho}'$ are transferred to 
$\hat{\rho}$, such as characteristic functions, quasi probability
distributions, etc. In particular, if  $\hat{\rho}'$ describes a Gaussian state,
also $\hat{\rho}$ corresponds to a Gaussian state.

A deeper study of the QAT applied to mixed states and how it can be used to
describe dissipation and decoherence analyzing the transformation properties of
master equations like the Lindblad one under the QAT is the object of work under
progress and will be presented elsewhere.

%
%

Another interesting point to study is whether the free particle states with
many humps are physically observable and measurable, since their harmonic
oscillator counterpart, the Hermite-Gauss and Laguerre-Gauss states are
``visible'' in Quantum Optics using lasers \cite{Fotonics}. Let us consider, for
instance, a two-hump wave packet $\phi_{\vec{a}}^{(1,0,\ldots,0)}(\vec{x},t)$ in
two or three dimensions with the humps in the transversal direction to that of
the mean velocity $\vec{v}_0$. The separation of the two maxima
of $|\phi_{\vec{a}}^{(1,0,\ldots,0)}|^2$ (see Fig. \ref{grafiquita}) is greater,
in a factor 1.6, than the uncertainty in position $\Delta x_1$. Therefore the
two humps should be measurable, and in fact, if this wave packet propagates
in a bubble or wire chamber, two parallel, divergent tracks would be observed
(if times $t<<\tau$ are considered). For a three-hump wave packet, the
separation among consecutive maxima (see Fig. \ref{grafiquita}) is smaller
than the uncertainty in position, although the distance between the more
separated maxima is greater than the uncertainty in position. This, together
with the fact that the central maximum is smaller than the external ones,
suggests that only two, overlapping thick tracks would be 
observed in a bubble or wire chamber. A similar behavior for a larger number 
of humps is expected. We think that these points deserve further study
and clarification.

The ideas developed here could also be applied to relativistic systems, 
particularly to the free particle in de-Sitter space-time, where the ordinary
formulation of quantum theory does not find a natural physical vacuum
\cite{desitter}. In this sense, the generalization of our approach to the
relativistic case would provide a hierarchy of states where the first state, the
relativistic counterpart of the Gaussian wave packet \cite{RHP,RHP2}, plays the
role of a vacuum \cite{relativistic-QAT}.


\section*{Acknowledgments}

Work partially supported by the Spanish MCYT, Junta de Andaluc\'\i a and
Fundaci\'on S\'eneca under projects FIS2008-06078-C03-01,
P06-FQM-01951 and 08816/PI/08. 

\section*{Appendix: Derivation of coherent, squeezed, number states for the free particle}
\label{Appendix}

In this Appendix we write down the configuration space wave functions for
coherent, squeezed, number states for the harmonic oscillator, and derive the
corresponding wave functions for the free particle using the QAT. Let us first
introduce some notation in order to simplify the expressions. Let
$\delta'=1+i\tan(\omega t')$, and $\delta_r'=1+i e^{2r}\tan(\omega t')$. These
expressions, under the classical Arnold transform (\ref{ArnoldClasico}), are
mapped to their corresponding unprimmed versions, $\delta=1+i\omega t$ and 
$\delta_r=1+i e^{2r}\omega t$.

Let us also define the dimensionless quantity
\begin{eqnarray}
q'&=&\frac{\sqrt{\frac{m\omega}{\hbar}} (x' -{x_0}\cos(\omega t') -
\frac{p_0}{m\omega} \sin(\omega t' ))}{(e^{2 r} \sin^2(\omega t')+e^{-2
r}\cos^2(\omega t'))^{1/2}}\nonumber \\ & & =\sqrt{\frac{m\omega}{\hbar}}
\frac{|\delta'|}{|\delta_r'|}
e^{r}(x' -{x_0}\cos(\omega t' )-\frac{p_0}{m\omega} \sin(\omega t' ))\,.
\end{eqnarray}

The configuration space wave functions are then written as: 
\begin{equation}
\psi'^n_{(a,r)}(x',t')= 
{\cal N}_n e^{r/2}\left(\frac{|\delta'|}{|\delta_r'|}\right)^{\frac{1}{2}}
\left(\frac{\delta_r'{}^*}{|\delta_r'|}\right)^{n+\frac{1}{2}}
 \hbox{\Large \it e}^{i\theta(x',t')}\hbox{\Large \it e}^{-q'^2/2} H_n(q') \,,
\label{squeezedstates}
\end{equation}
where the phase $\theta'(x',t')$ is given by:
\begin{eqnarray}
\theta'(x',t')&=& \frac{1}{|\delta'|^2}
\biggl[\frac{1}{2}
\left(\frac{1}{m\hbar \omega}p_0^2-\frac{m \omega }{\hbar}x_0^2\right) 
\tan (\omega t') + \frac{1}{\hbar} p_0 x_0  \nonumber \\
& & +
e^{-r}|\delta_r'|  \biggl(\frac{p_0}{\sqrt{m\hbar\omega}} 
-\sqrt{\frac{m\omega}{\hbar}} x_0\tan (\omega t')\biggr)q' \nonumber \\
& & + \sinh(2 r)  \tan(\omega t')q'{}^2\biggr]\,.
\end{eqnarray}

To compute the QAT of these states, $\psi_{(a,r)}^n(x,t)$, it is useful to
write down a ``dictionary'', completing (\ref{ArnoldClasico}):
\begin{equation}
\begin{array}{rclcrcl}
\omega t'& \rightarrow &\tan^{-1}(\omega t)  
& & 
x'&\rightarrow &\frac{x}{|\delta|}      
\\
\cos(\omega t')& \rightarrow & \frac{1}{|\delta|} 
& & 
\sin(\omega t')& \rightarrow & \frac{\omega t}{|\delta|}
\\
\delta'& \rightarrow & \delta 
& &  
\delta_r'& \rightarrow & \delta_r 
\\
 \psi' &\rightarrow & \psi =
\frac{1}{\sqrt{|\delta|}}e^{i\omega t \frac{x^2}{4L^2|\delta|^2}}\psi' \,.
& & & &
\end{array} 
\end{equation}

Using this dictionary, the adimensional quantity $q'$ changes to: 
\begin{equation}
q'\rightarrow q= \frac{x -x_0+\frac{p_0}{m} t }{\sqrt{2}Le^{-r}|\delta_r|}\,,
\end{equation}
and the phase $\theta(x',t')$ changes to:
\begin{eqnarray}
\theta'(x',t') \rightarrow \theta(x,t)
&=& \frac{1}{|\delta|^2}\biggl[\frac{1}{2}
\left(\frac{1}{m\hbar \omega}p_0^2-\frac{m \omega }{\hbar }x_0^2\right) \omega t
+\frac{1}{\hbar} p_0 x_0  \nonumber 
\\
& & +
e^{-r}|\delta_r|  \biggl(\frac{p_0}{\sqrt{m\hbar\omega}} 
-\sqrt{\frac{m\omega}{\hbar}} x_0 \omega t\biggr)q \nonumber 
\\
& & + 
\sinh(2 r) \omega t q{}^2\biggr]\,.
\label{free-theta}
\end{eqnarray}

Therefore, the corresponding coherent, squeezed, number state for the free
particle reads:
\begin{eqnarray}
 \psi_{(a,r)}^n(x,t)=  {\cal N}_n e^{r/2}\frac{1}{\sqrt{|\delta_r|}}
\left(\frac{\delta_r{}^*}{|\delta_r|}\right)^{n+\frac{1}{2}}
 \hbox{\Large \it e}^{i\omega t \frac{x^2}{4L^2|\delta|^2}}
\hbox{\Large \it e}^{i\theta(x,t)}\hbox{\Large \it e}^{-q^2/2} H_n(q)\,.
\label{squeezedstates-freeparticle}
\end{eqnarray}

\end{document}